\documentclass[superscriptaddress,aps,twocolumn,showpacs,showkeys,notitlepage,amsmath, amssymb,
]
{revtex4-1}
\usepackage{graphicx}
\usepackage[mathlines]{lineno}
\usepackage[svgnames]{xcolor}
\usepackage{hyperref}
\usepackage{color}
\hypersetup{
	colorlinks = true,
	linkcolor = Blue,
	citecolor = Blue,
	urlcolor  = Blue
}

\newcommand{\bco}{Ba$_2$CoO$_4$}
\newcommand{\co}{CoO$_4$}

\usepackage{comment}
\usepackage{ltablex}
\usepackage{threeparttable}
\usepackage[referable]{threeparttablex}
\usepackage[caption=false]{subfig}
\newcommand{\phantomsubfloat}[1]{
    {
        \captionsetup[subfigure]{labelformat=empty}
        \subfloat[][]{#1}
    }%
}
\usepackage{siunitx}

\usepackage{pbox}
\usepackage{booktabs,microtype,afterpage} 

\begin{document}

\title{Frustrated network of indirect exchange paths between tetrahedrally coordinated Co in \bco{}}
\author{Ifeanyi John Onuorah}
\email[]{ifeanyijohn.onuorah@unipr.it}
\affiliation{Department of Mathematical, Physical and Computer Sciences, University of Parma, Italy}
\author{Muhammad Maikudi Isah}
\affiliation{Department of Mathematical, Physical and Computer Sciences, University of Parma, Italy}
\author{Roberto De Renzi}
\affiliation{Department of Mathematical, Physical and Computer Sciences, University of Parma, Italy}
\author{Pietro Bonf\`{a}}
\affiliation{Department of Mathematical, Physical and Computer Sciences, University of Parma, Italy}

\date{\today}

\begin{abstract}
 
We present a detailed study of the electronic and magnetic interactions of \bco, structurally very uncommon because of the isolated \co{} distorted tetrahedral coordination. We show the presence of Co($d$)$-$O($p$) hybridized states characterized by  spin polarized oxygen atoms, with their magnetic moments parallel to that on Co. The calculated isotropic exchange interaction parameters, which include the contributions from ligand spins, demonstrate the presence of a 3D network of magnetic couplings, that are partially frustrated in the identified magnetic ground state. 
Our results indicate that the dominant indirect exchange mechanism responsible for this ground state is mediated by O atoms along the Co$-$O$\dotsb$O$-$Co path.
\end{abstract}

\maketitle
\section{\label{sec:intro} Introduction}
Cobalt based oxides display unique spin states, electronic properties and magnetic interactions arising from the nature of the coordinating oxygen atoms and the interaction of the existing multiple degrees of freedom. \bco{} is one of the few known cobalt oxides that exhibits tetrahedral coordination of all Co magnetic ions with a rare case of indirectly linked tetrahedra, as opposed to both the corner and side sharing ones. Indeed, octahedrally coordinated Co has been the subject of more extensive research~\cite{goodenough1971,potze1995,korotin1996,felser1999,hua2005,sugiyama2006,sugiyama2006,pardo2007,nozaki2007,kunes2012} owing to the potential similarities with the cuprate superconductors~\cite{takada2003,matsuno2004}. Conversely, oxides with tetrahedrally coordinated Co are less studied because only but a few are found to be stable in this configuration~\cite{zhang2012}. \bco{} is one of them, and it offers the opportunity to study the interplay between lattice, charge, and spin degrees of freedom for Co in the tetrahedral environment.

\begin{figure*}
\centering
\includegraphics[width=17.0cm]{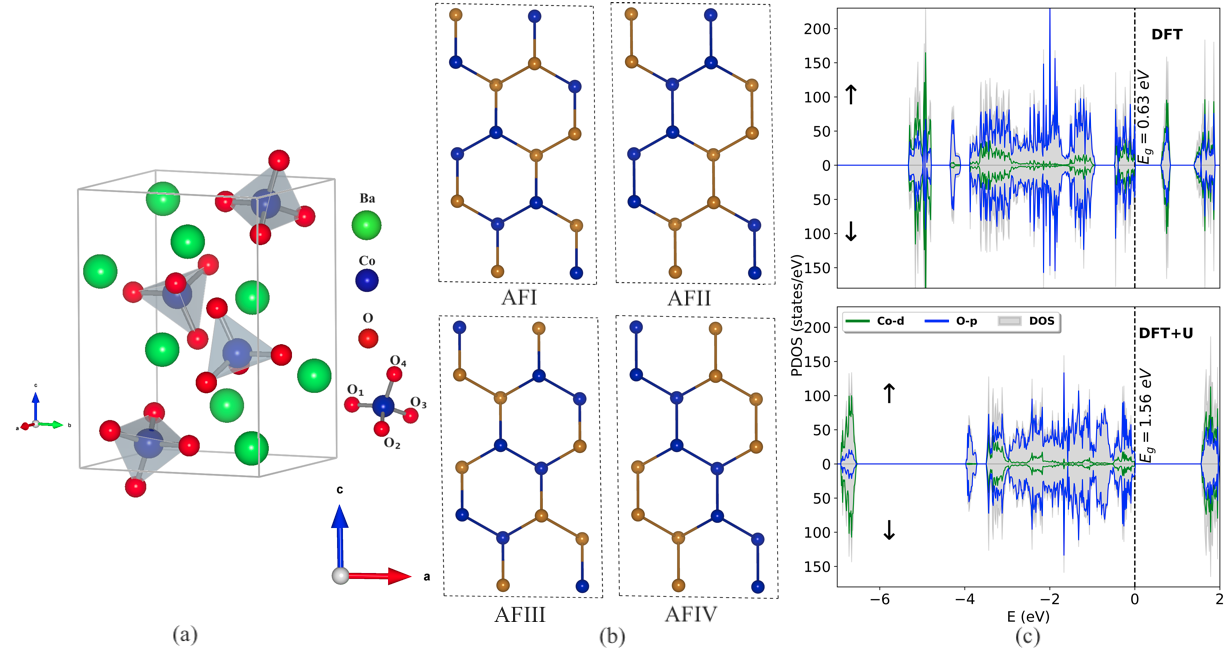}
\phantomsubfloat{\label{fig:figure1a}}
 \phantomsubfloat{\label{fig:figure1b}}
 \phantomsubfloat{\label{fig:figure1c}}
\vspace{-2\baselineskip}
\centering
\caption[Unitcell ]{\label{fig:figure1}(a) Unit cell of  \bco{} showing the isolated tetrahedral coordination of the Co (blue spheres) and O (red spheres) atoms, the green spheres are the Barium atoms. (b) Representation of the four collinear magnetic AF structures considered in the magnetic unit cell. The FM structure is not shown. For clarity, only the Co atoms are shown, color blue for  spin up and brown for spin down polarization. Notice that the $ac$ plane is buckled and the plotted Co atoms are not at the same $b$ coordinate. (c) The density of states (grey background) and projected density of states (Co-$d$ and O-$p$) for both the majority spin ($\uparrow$) and minority spin ($\downarrow$) channel of the AFI structure (for DFT and  DFT+U calculations). The zero energy is set to the valence band maximum.}
\end{figure*}

\bco{} crystallizes in a monoclinic lattice structure having space group P2$_1$/n (no. 14) and lattice parameters $a$ = 5.9176  \AA, $b$ =7.6192, $c$ = 10.3970 \AA{} and $\beta$= 91.734\si{\degree} ~\cite{jin2006}.  The unit cell contains four distorted  CoO$_{4}$ tetrahedra with tetrahedral angles ranging from 104.47\si{\degree} to 112.87\si{\degree} ~\cite{jin2006}. Each tetrahedron complex is isolated from the other (See Fig.~\ref{fig:figure1a}), with Co$\dotsb$Co  and O$\dotsb$O distances above 4.7 \AA{} and  3.0 \AA{} respectively ( where $\dotsb$ indicates the distance between atoms across neighbor tetrahedra). From this geometric inspection, one may expect extremely weak exchange couplings and, as a consequence, very low magnetic transition temperatures. Nonetheless, a surprisingly high Curie-Weiss temperature parameter $|\Theta|$ $\approx$ 110 K and antiferromagnetic order below $T_{N}=25$~K are observed~\cite{jin2006}. To explain the large difference between  $|\Theta|$ and  $T_{N}$, magnetic frustration has been suggested~\cite{jin2006,koo2006}. Also, spin dimer analysis with tight binding calculation  attributes the origin of the difference in temperature to layered magnetic frustration \cite{koo2006}, while an alternative analysis of the spin dynamics with inelastic neutron scattering~\cite{zhang2019}  and a very recent DFT study~\cite{zhang2021} assert that quasi 2D-magnetism could be realized in non-layered \bco{}.  Hence, there is the need to fully understand the magnetic interactions together with the nature and origin of the probable indirect exchange interaction that drives magnetism. Also, multiple experimental results establish that the magnetic structure has propagation vector \textbf{k} = (0.5,0,0.5), but the actual alignment of the magnetic moments localized on Co is still debated~\cite{boulahya2006,russo2009,zhang2019,ryu2019}. Moreover, the realization of an intermediate spin state for Co was recently proposed to be the cause of the reduced magnetic moment observed experimentally~\cite{zhang2019}.  All these points are indeed open-ended issues that demand further theoretical investigation.

For these reasons, we present here a thorough investigation of the electronic properties and magnetic interactions of \bco{} using first principles calculations.  We compare the stability of the proposed magnetic structures, discuss the roles of oxygen and the isolated distorted tetrahedra in the system. We also calculate the isotropic contribution to the exchange coupling parameters. The obtained values justify many of the magnetic properties highlighted above.

\section{\label{sec:method}Method}
 Ab initio simulations were performed using Density Functional Theory (DFT) within the projector augmented pseudopotentials ~\cite{blochl1994} and the GGA for the exchange correlation functional (Perdew-Burke-Ernzerhof ~\cite{pbe1996}) as implemented in the Quantum Espresso code~\cite{qe2009}. We have also considered the role of electron-electron interaction on Co$-d$ orbitals by setting the effective Coulomb interaction $U_{\mathrm{eff}} = (U - J)$ value to 5.75 eV, obtained self consistently in the DFT + U scheme \cite{solovyev1996,dudarev1998,cococcioni2005,kulik2006,timrov2018}. The magnetic structure with the experimental propagation vector (0.5, 0, 0.5) was adopted with the use of the  2$\times$1$\times$2 supercell  consisting of 112 atoms. All our calculations were carried out within the collinear spin formalism, although the reported AF magnetic structure is non-collinear with a main component either along the $c$~\cite{boulahya2006} or the $a$ axis~\cite{russo2009,zhang2019,ryu2019}.  The cut off used for the plane waves and the charge density are 100 Ry and 900 Ry respectively, with the Brillouin zone sampled using a 3$\times$5$\times$1 mesh of K-points~\cite{kpoint1976}. The Marzari-Vanderbilt~\cite{mv1999} smearing with width of 0.005 Ry was used except for the plots and projection of the density of states where the optimized tetrahedron method~\cite{mitsuaki2014} was adopted. Atomic positions were optimized to force and energy thresholds of 1$\times$10$^{-3}$ a.u and 1$\times$10$^{-4}$ Ry respectively while the experimental lattice parameters~\cite{jin2006} were adopted. We represented the spin polarized bands in the Wannier function basis~\cite{mostofl2008,pizzi2020} by projecting onto Co 3$d$ and O 2$p$ orbitals that span the subspace of 272 Bloch bands.

\footnotetext{See Supplemental Material at [below], which includes Refs.~\cite{momma2011,heisenberg1928,liechtenstein1987,mazurenko2005,mazurenko2007,pchelkina2014,solyom2007} for (i) Magnetic symmetry and structure optimization (ii) Search for occupation of Co spin states(iii) Projected Density of States (iv) Details of Wannier Functions and exchange coupling
(v) Oxygen spin contribution to the magnetic interaction.
(vi) Magnetic structure stability with the Heisenberg Model
(vii) Mean-Field Estimation of T$_N$ and $\Theta$}

\section{\label{sec:Results}Results and Discussion}
 \subsection{\label{sec:Magsym} Magnetic order}
First, we consider the stability of the different magnetic structures. Fig~\ref{fig:figure1b} shows the four AF magnetic orders in the collinear formalism obtained from the maximal subgroups of the magnetic space groups that allow non-zero magnetic moment on Co, consistent with the experimental propagation vector (0.5, 0, 0.5)~\cite{mato2015} (see Table S1 in the supplemental material (SM)~\cite{note1}). Limiting us to the cases with spin polarization along the $a$ or the $c$ axis, the magnetic symmetry distinguishes four {collinear} AF configurations labelled AFI, AFII, AFIII and AFIV structures. Notably, several of the quoted experiments~\cite{ryu2019,zhang2019,boulahya2006,russo2009} report different magnetic structures (non-collinear versions of AF structures reported above). Additionally, we considered the ferromagnetic structure labelled FM and the other AF structures discussed in Sec.~I of the SM~\cite{note1}, all having higher energies. 

\begin{table}[!h]
    \centering
    \caption{Total energy differences with respect to AFI are presented for both DFT+U and DFT optimized atomic positions in meV/f.u. 
    The moments on Co (obtained from the projection on Co atomic orbitals) for the four AF structures and the FM structure described in the text are presented.}
    \label{tab:table1}
    \begin{ruledtabular}
    \begin{tabular}{c c c c c}
         Structure &  $\Delta{E}^{DFT+U}$ &  $\Delta{E}^{DFT}$ & m$_{\mathrm{Co}}^{DFT+U}$   &  m$_{\mathrm{Co}}^{DFT}$\\
         & [meV/f.u.] & [meV/f.u.] & [$\mu_\mathrm{B}$]  &  [$\mu_\mathrm{B}$]\\
          \hline
         AFI   & 0.0 & 0.0     & 2.87 & 2.82  \\
         AFII  & 0.21 & 2.16   & 2.87  & 2.82  \\
         AFIII & 1.64 & 4.74   & 2.87  & 2.82  \\
         AFIV & 3.83 & 8.37   & 2.87  & 2.82  \\
         FM    & 32.17 & 44.43 & 2.88 & 2.84 \\
    \end{tabular}
    \end{ruledtabular}
\end{table}

Table~\ref{tab:table1} shows that our DFT+U (DFT) results for the AFI  structure at the relaxed atomic positions have the lowest energy with respect to the AFII, AFIII, AFIV and FM structures, separated by  0.208 (2.164), 1.640 (4.744), 3.830 (8.365) and 32.169 (44.428)  meV/f.u.  respectively. The \emph{very} small energy differences between the AF structures suggest that the magnetic order results from the competition of the various magnetic interactions present in the system. 

Table~\ref{tab:table1} also shows that, for all these structures, both DFT+U and DFT determination of the magnetic moment values on each Co remain in good agreement with experimental values: 2.69(4)~$\mu_\mathrm{B}$~\cite{zhang2019}, 3.5~$\mu_\mathrm{B}$~\cite{ryu2019} and  3.23~$\mu_\mathrm{B}$~ \cite{boulahya2006}. In addition an induced moment of  $\approx$ 0.5~$\mu_\mathrm{B}$, parallel to that on Co,  is found on each O atoms.  This moment on O is common to all the considered magnetic structures and its value does not change significantly among them. From now on we quote only results related to the AFI ground state structure.

\subsection{\label{sec:Elecstruct} Hybridization effects}
Fig.~\ref{fig:figure1c} shows the density of states (DOS) and projected density of states (PDOS) with atomic orbital contributions from the Co$-d$ and O$-p$. The plot shows that the band gap increases from 0.63 eV with DFT to 1.56 eV with DFT+U, indicating the relevance of electron (Coulomb) correlation in the system. These small bandgap values suggest a semiconducting behaviour for \bco. Strong hybridization between Co$-d$ and O$-p$ orbitals is indicated by their large contributions to the PDOS both in the valence and in the conduction states around the Fermi energy (Fig.~\ref{fig:figure1c}).

This  strong hybridization is indeed the cause of the relatively large spin polarization on O. Our results are in agreement with very recent DFT calculations \cite{zhang2021} and with a model supported by DFT \cite{lin2021} that reveals how the delocalization of O electrons on the Co sites is a consequence of the kinetic energy optimization.

\begin{table}[!h]
 \begin{threeparttable}
\caption{\label{tab:table2} Occupancy of the Co$-d$ states and an average representative O state (distorted tetrahedral) obtained with DFT+U.}
\begin{ruledtabular}
 \begin{tabular}{ l @{\hspace{0.8em}} l @{\hspace{0.8em}} l @{\hspace{0.8em}} l @{\hspace{0.8em}} l@{\hspace{0.8em}} l@{\hspace{0.8em}} l@{\hspace{0.8em}} l l} 
                          Spin state    & d$_{x^2-y^2}$      &  d$_{z^2}$      &  d$_{xz}$   &  d$_{yz}$     & d$_{xy}$    &  total & total\\ 
                                                   &       &        &     &      &     &  Co-$d$ & O-$p$~\tnotex{tnxx86}\\ 
     \colrule
    majority      &       1.00  &   1.00 &  0.99 &  0.99 & 0.99  & 4.97 & 2.90\\
     minority    &       0.61  &   0.60 & 0.57  & 0.36 & 0.33  & 2.47 & 2.51 \\   
     total   &      &             &             &           &            &7.44 & 5.41
       \end{tabular} 
  \end{ruledtabular}
  \begin{tablenotes}
			\item[a]  \label{tnxx86} For detailed O occupancies see Table S5 in SM~\cite{note1}. 
		\end{tablenotes} 
 \end{threeparttable}
  \
\end{table}

Table \ref{tab:table2} reports the occupation numbers of the Co$-d$ and O$-p$ levels (more details in Tables S4 and S5 in the SM~\cite{note1} ). Notice that these occupation numbers are independent of the AF structures. From a purely ionic picture, the high spin state of Co$^{4+}$ is described as a $e_g^2 t_{2g}^3$ configuration with the maximum value of spin S=5/2. However, the hybridization of Co and O states breaks  this simplistic description, and the occupation of Co$-d$ levels is found to be substantially higher, approaching $\sim 7.5$ electrons (see Tab.~\ref{tab:table2}). As a consequence, the magnetic moment on Co atoms is only about 2.87~$\mu_\mathrm{B}$. 
This is actually the result of the predominant localization of extra holes on oxygen sites, as shown in Tab.~\ref{tab:table2}, an effect also responsible for the observed magnetic polarization of 0.5 $\mu_{\mathrm B}$ on each oxygen. Indeed, each Co in a tetrahedron couples with four ligands that totally contribute  $\approx$ 2 hole states to its $d$ orbitals.

A proper description of magnetism in the presence of this strong $p$-$d$ hybridization requires the inclusion of the ligand states. The overall magnetic moment of this entity is $\approx${5}~$\mu_\mathrm{B}$ and it behaves as an effective total spin $5/2$, in agreement with susceptibility measurements~\cite{candela1973,boulahya2006,jin2006}. The distortion of the tetrahedron completely lifts the degeneracy of the $d$ levels, albeit with a reduced crystal field splitting, typical of tetrahedral vs. octahedral coordination. This results in a fine adjustment of the fractional occupation of the orbitals, as reported in Table~S4 and S5 of the SM~\cite{note1}.

\begin{figure}[!h]
\includegraphics[width=8.2cm]{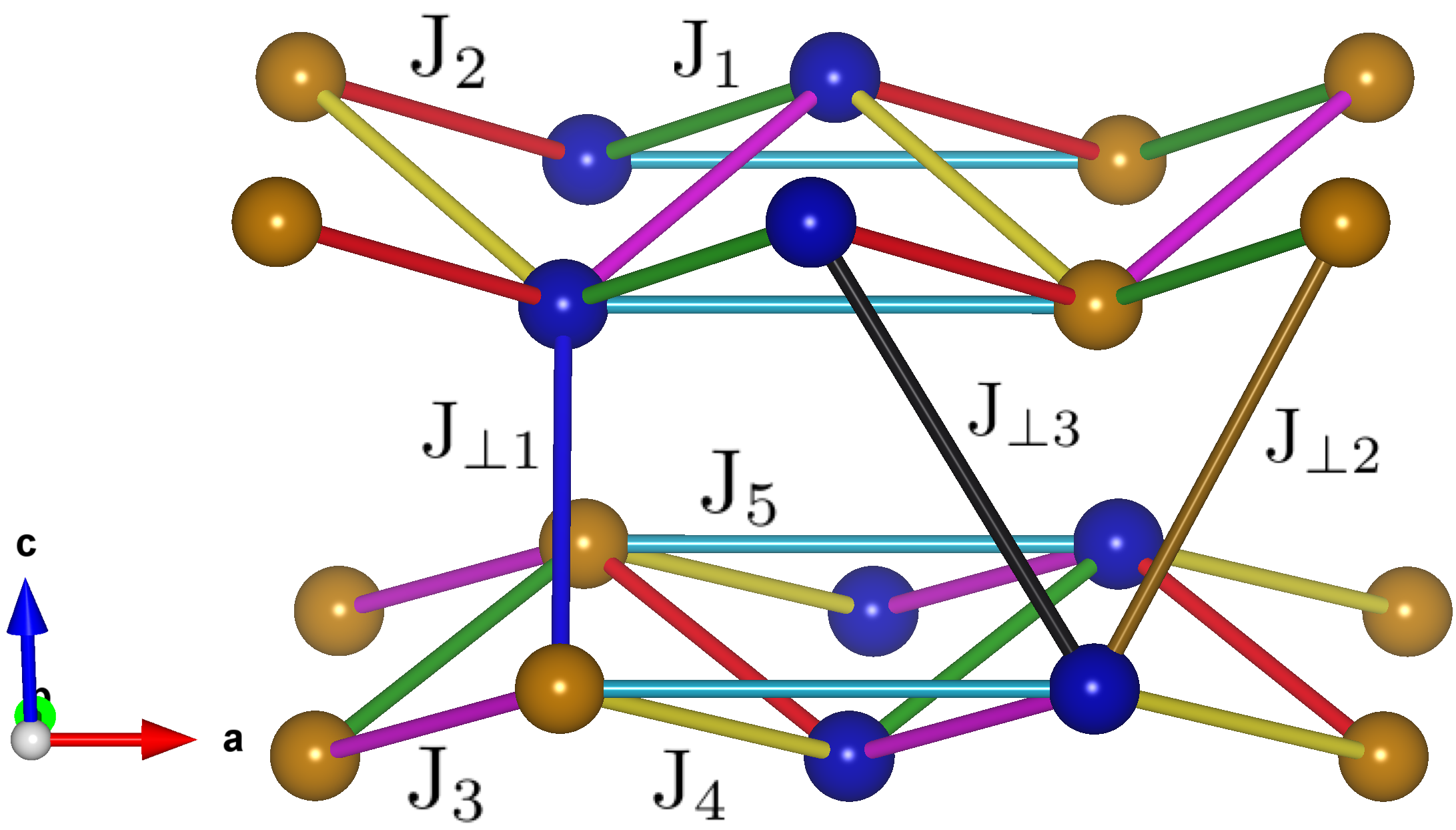}
\centering
\caption[Untc ]{\label{fig:figure2}Magnetic structure of AFI  (blue spheres  represent $\uparrow$ while brown spheres represent $\downarrow$  on Co) showing the eight exchange couplings considered in this work: five intralayer $J_i$ and three interlayer $J_{\bot i}$. For clarity the couplings are color coded and only one of each is labelled.} 
\end{figure}  
 
\begin{table*}[!hbtp]
 \begin{threeparttable}
\centering
\caption[exchange coupling]{\label{tab:table3} Calculated isotropic exchange constants  (in meV) between pairs of Co spins denoted as $J^{\mathrm{Co} \dotsb \mathrm{Co}}$ and those including the effects of the ligand spins denoted as $J^{\mathrm {eff}}$ obtained with both DFT and DFT+U.  The  DFT+U relaxed  distances between two Co (d$_{\mathrm{Co} \dotsb \mathrm{Co}}$) (in \AA), distances between two O atoms in adjacent tetrahedron (d$_{\mathrm{O} \dotsb \mathrm{O}}$) (in \AA)  and the corresponding  Co$-$O$\dotsb$O, O $\dotsb$O$-$Co  angles ($\angle$) (in \si{\degree}) that bridge the Co$\dotsb$Co exchange interactions are shown for each coupling.}
\begin{ruledtabular}
 \begin{tabular}{ l@{\hspace{0.6em}} p{1.2cm}  p{1.0cm} p{2.3cm} p{1.5cm} p{1.5cm}  p{1.5cm}  p{1.5cm}}
 &&&&\multicolumn{2}{c}{with DFT} & \multicolumn{2}{c}{with DFT+U}\\
          \cline{5-6}\cline{7-8}\\ 		        
       & d$_{\mathrm{Co} \dotsb \mathrm{Co}}$ & d$_{O \dotsb O}$ & \hfil $\angle$ &$J^{\mathrm{Co} \dotsb \mathrm{Co}}$ & $J^{\mathrm {eff}}$&$J^{\mathrm{Co} \dotsb \mathrm{Co}}$ &$J^{\mathrm{eff}}$ \\
       \colrule
    $J_1$                           & 4.756    &  3.577  &83.15, 83.15  & -0.350(1)   & -0.149(1)    & -0.241(1)   & 0.024(1)    \\
    $J_2$                           & 4.865    &  3.245  &90.61, 90.61     &-0.496(2) & -0.241(3)    &  -0.337(1)  & -0.040(1)   \\ 
    $J_3$                           & 5.364    &  3.340  &129.92, 114.64  & -2.069(6)& -2.941(10)  &  -1.902(2)  & -2.468(3)    \\ 
    $J_4$                           & 5.461    &  3.273  &137.55, 115.88  & -3.031(3)& -4.478(5)    &  -2.723(3)   & -3.654(4)   \\ 
    $J_5$                          & 5.918     &  3.004  &145.68, 139.09  & -2.268(9)& -3.125(14)  &  -2.334(2)   & -3.094(1)    \\ 
    $J_{\bot 1}$                 & 5.192     &  3.208  &119.43, 104.97  & -1.219(10)&-1.489(13)  &  -1.016(1) & -1.032(1)    \\ 
    $J_{\bot 2}$                 &5.917      &  3.289  &135.40, 115.55  & -0.793(1)  & -0.983(3)   &  -0.757(1)  & -0.866(1)    \\ 
    $J_{\bot 3}$                 &6.072      &  3.304  &141.64, 123.41  & -0.754(4)  & -0.939(6)   &  -0.719(1)  & -0.856(2)     \\  
    \end{tabular}
\end{ruledtabular}
  \end{threeparttable}
  	\
\end{table*}

\subsection{\label{sec:magexhange}Exchange coupling constants and interaction mechanism }

We can now discuss the Co$\dotsb$Co exchange coupling interactions in \bco{} together with the contributions from the induced spin at the ligand sites. The ligand spins, as we have seen, are non-negligible and  their effects on the exchange interactions have been discussed in a number of publications on different systems~\cite{logemann2017,besbes2019,solovyev2021}.   Starting with the Heisenberg Hamiltonian, 
\begin{equation}
H=- \sum_{i\neq j}{ J_{ij} \mathbf{S}_i  \cdot  \mathbf{S}_j}, \label{eq:heismain}
\end{equation}
we treat both the spins on Co and O as localized, where  $J_{ij}$ is the isotropic exchange constants between spin $\mathbf{S}$ (normalized to 1) at sites i and j. The isotropic exchange constants were calculated by the Green’s function approach~\cite{korotin2015,xu2021} with the projected Wannier functions as the localized basis. 

Starting from the exchange constant contribution from interacting Co pairs $J^{\mathrm{Co} \dotsb {\mathrm{Co}}}$, we use the downfolding procedure by Solovyev~\cite{solovyev2021} as a means to include the effects of the ligand spins (in our case between 
${\mathrm{Co}} \dotsb {\mathrm{O}}$ and ${\mathrm{O}} \dotsb {\mathrm{O}}$ pairs) transferring them to effective coupling interactions between spins at Co sites. This results in the effective exchange constants $J^{\mathrm{eff}}$. 

\subsubsection{Co$\dotsb$Co interaction and ligand spins contribution}
Figure \ref{fig:figure2} shows the spin alignment between Co$\dotsb$Co pairs, which can be described as a layered, buckled structure of  dimer chains along the $a-b$ axis, with alternate ordering of  the spins in adjacent layers along the c axis. Following the notation used in Ref~\cite{zhang2019}, we can define two groups of exchange couplings, the intralayer coupling parameters ($J_1$, $J_2$, $J_3$, $J_4$ and $J_5$, within the $ab$ plane)  and the inter-layer coupling parameters ($J_{\bot 1}$, $J_{\bot 2}$ and $J_{\bot 3}$, along the $c$ axis). 

The calculated exchange coupling constants for both $J^{\mathrm{Co} \dotsb \mathrm{Co}}$ and $J^{\mathrm {eff}}$, for distances ranging from 4.76 \AA{} to 6.07 \AA{}, are reported in Table~\ref{tab:table3}. The magnitude of the exchange interactions between pairs of Co atoms farther apart are very small and vanishing relative to the reported values. It is surprising that $J_1$ and $J_2$ have the weakest magnitude of the reported exchange parameters, despite having the shortest Co$ \dotsb$Co distances.
All the exchange parameters for the Co$\dotsb$Co contribution are antiferromagnetic ($J<$0) for both DFT and DFT+U calculations, including $J_1$, $J_3$ and $J_{\bot 3}$ which couple  parallel spins in the AFI magnetic ground state (see Fig.~\ref{fig:figure2}). 

The magnetic structure is  stabilized within the $a-b$ plane by the stronger $J_4$ and $J_5$ interactions. A wave vector dependent Heisenberg energy calculation with the DFT+U exchange constants confirms that the lowest energy state is at the (0.5,0,0.5) propagation vector (see sec.~VI of the SM~\cite{note1}), in agreement with experimental findings, and that the lowest energy state is AFI, in agreement with collinear DFT+U  and DFT calculations.

The  inclusion of the ligand contributions described by the $J^{\mathrm {eff}}$ model further weakens $J_1$ and $J_2$ and strengthens the other dominant exchange couplings in both DFT and DFT+U calculations. The above description includes the presence of frustrated antiferromagnetic interactions, as it was previously suggested~\cite{boulahya2006,koo2006}. Indeed, frustration justifies the difference in the mean field estimates of $T_N^m$ and $\Theta^m$ that yield $T_N^m$ = 41.66 K and $|\Theta^m|$=80.45 K using  $J^{\mathrm{Co} \dotsb \mathrm{Co}}$, and  $T_N^m$ = 51.57 K and $|\Theta^m|$=96.58 K using  $J^{\mathrm{eff}}$ in qualitative agreement with the experimental values  of  $T_N$ = 25K and $|\Theta|$=110K (using results from DFT+U simulations and assuming S=5/2, see sec.~VII of the SM~\cite{note1} for additional details).

Further, the exchange coupling values reported in Table~\ref{tab:table3} show that the inter-layer couplings $J_{\bot}$ along the $c$ axis are non-negligible, indicating that the magnetic order has a 3D character, ruling out  the earlier reported quasi-2D nature~\cite{zhang2019}. The fact that $J_\perp$ do not vanish is also evidenced by the total energy differences between the AFI and AFII structures which have identical spin alignment within the $ab$ layers but differ in the stacking along $c$ (see  Fig.~\ref{fig:figure1b}).

In addition, treating both Co and O moments as localized spins allows us to also quantify the strong magnetic interaction within each tetrahedron complex. The Co$-$O couplings are all ferromagnetic with very large values  ranging from 48.7 to 59.4 meV.  This is due to the strong Co($d$)-O($p$) hybridization and justifies the intuitive understanding of the tetrehedron magnetic unit, with O spins parallel to that of the central Co, as an effective moment even in the high temperature paramagnetic regime. This unit has a total moment of $4.88~ \mu_{\mathrm B}$ obtained by summing the moments from the projection on Co and O atomic orbitals.

\begin{figure}[!h]
\includegraphics[width=8.5cm]{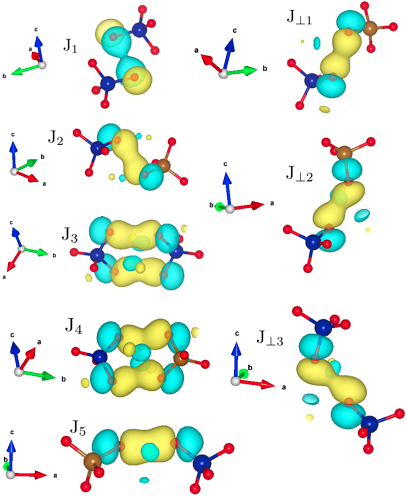}
\centering
\caption[Wannier functions]{\label{fig:figure3} Isosurface plots of the Wannier functions (spin-up) showing the $p$ orbitals overlap between the O$\dotsb$O atoms along the paths that mediate the exchange interaction between Co in neighbour tetrahedra. Positive isosurface is rendered in yellow and negative in blue, all with the same iso-level.}  
\end{figure}

\subsubsection{Exchange mechanism}
Let us address the exchange mechanism acting
among adjacent \co{} tetrahedra. The lack of hybridization between Ba-$p$ states and O-$p$ or Co-$d$ orbitals makes the contribution of Ba to exchange interactions negligible, even though  Co$-$O$\dotsb$Ba$\dotsb$O$-$Co is the shortest path between adjacent Co atoms \cite{katukuri2020}. Unsurprisingly, the inclusion of Ba-p states in the Wannier function basis does not modify the calculated exchange couplings.

The Wannier 3$d$ orbitals centered at the Co sites have large “tails" spreading to the coordinating oxygen sites (see SM~\cite{note1} Fig. S4) suggesting a more significant role of oxygen in the exchange interaction. The exchange mechanism is clarified with an alternative Wannier basis that includes only the Co$-d$ orbitals, neglecting the O$-p$ orbitals, which results in vanishing values for the coupling parameters. These considerations support the view that exchange interaction via the Co$-$O$\dotsb$O$-$Co path dominates the inter-site magnetic interactions.

Indeed, a substantial overlap is found between oxygen orbitals of adjacent tetrahedron complexes, as shown in Fig~\ref{fig:figure3}.
The O$\dotsb$O distances and O$\dotsb$O$-$Co angles of the bonds that mediate the exchange interaction for each of the Co$\dotsb$Co exchange interaction paths are listed in Table~\ref{tab:table3}. The shortest Co$\dotsb$Co distances correspond to the weakest couplings $J_1$, $J_2$. Qualitatively, this is due to the fact that the O$\dotsb$O distance along these paths are the largest and the angles closest to 90 degrees. Table~\ref{tab:table3} shows that for these two weaker couplings the difference between the $J^{\mathrm{Co} \dotsb \mathrm{Co}}$ and $J^{\mathrm{eff}}$ calculations,  neglecting and including the O contribution respectively,  is small, in the order of 0.2 meV,  suggesting that the specific presence of a moment on O does not alter drastically the picture of the magnetic interaction, which is already captured by $J^{\mathrm{Co} \dotsb \mathrm{Co}}$. This is true also for the other couplings, where the relative difference between the $J^{\mathrm {eff}}$ and  $J^{\mathrm{Co} \dotsb \mathrm{Co}}$ calculations never exceeds 32\%.  

\section{\label{sec:concl}Conclusions}
In summary, using spin polarized DFT and DFT+U calculations we have determine the arrangement of magnetic moments in the crystallographic unit cell and we identified the important roles played by O anions in the structurally rare \bco, where the interesting ground state properties result from the effects of the Co($d$)-O($p$) hybridized states. We find that the oxygen-hole ligand contributions supersede the original debate on Co spin state, since the bound ligand holes account for the substantially reduced moment observed on Co. The $d$ level states are characterized by a broken tetrahedral symmetry which is observed in our DFT results by distorted Co$-$O bonds. The calculated exchange coupling constants reveal the 3D nature of the magnetic ordering and show that the not so large difference between the N\'eel and Curie-Weiss temperatures can be attributed to residual frustration. Therefore our results  show that the complex magnetic behavior of \bco{} is a consequence of the surprisingly strong but frustrated interaction between \co{} tetrahedra mediated by O$-p$ orbitals.

\begin{acknowledgments}
We thank Giuseppe Allodi and Paolo Santini for insightful discussions. We acknowledge funding from the SUPER (Supercomputing Unified Platform - Emilia-Romagna) regional project. We also acknowledge computing resources provided by  CINECA under  Project ID IsC58 and SUPER, the STFC scientific computing department’s SCARF cluster and the HPC resources at the University of Parma, Italy. 
\end{acknowledgments}

\bibliography{references}

\end{document}


\title{Supplemental Material for ``Frustrated network of indirect exchange paths between tetrahedrally coordinated Co in \bco{}"}
\author{Ifeanyi John Onuorah}
\email[]{ifeanyijohn.onuorah@unipr.it}
\affiliation{Department of Mathematical, Physical and Computer Sciences, University of Parma, Italy}
\author{Muhammad Maikudi Isah}
\affiliation{Department of Mathematical, Physical and Computer Sciences, University of Parma, Italy}
\author{Roberto De Renzi}
\affiliation{Department of Mathematical, Physical and Computer Sciences, University of Parma, Italy}
\author{Pietro Bonf\`{a}}
\affiliation{Department of Mathematical, Physical and Computer Sciences, University of Parma, Italy}

\date{\today}

\maketitle

\section{\label{sec:appendixoptmisedpos}  Magnetic symmetry and Structure Optimization}

Four structures from the maximal subgroups of the magnetic space groups that allow non-zero magnetic moment on Co with the experimental propagation vector (0.5, 0, 0.5)~\cite{mato2015} are considered and presented in Table~\ref{tab:SMtable1}.  With spin polarization along the $c$ axis, the magnetic symmetry results in 4 AF configurations labelled AFI, AFII, AFIII and AFIV.  

\begin{figure}[!h]
\centering
\includegraphics[width=15.5cm]{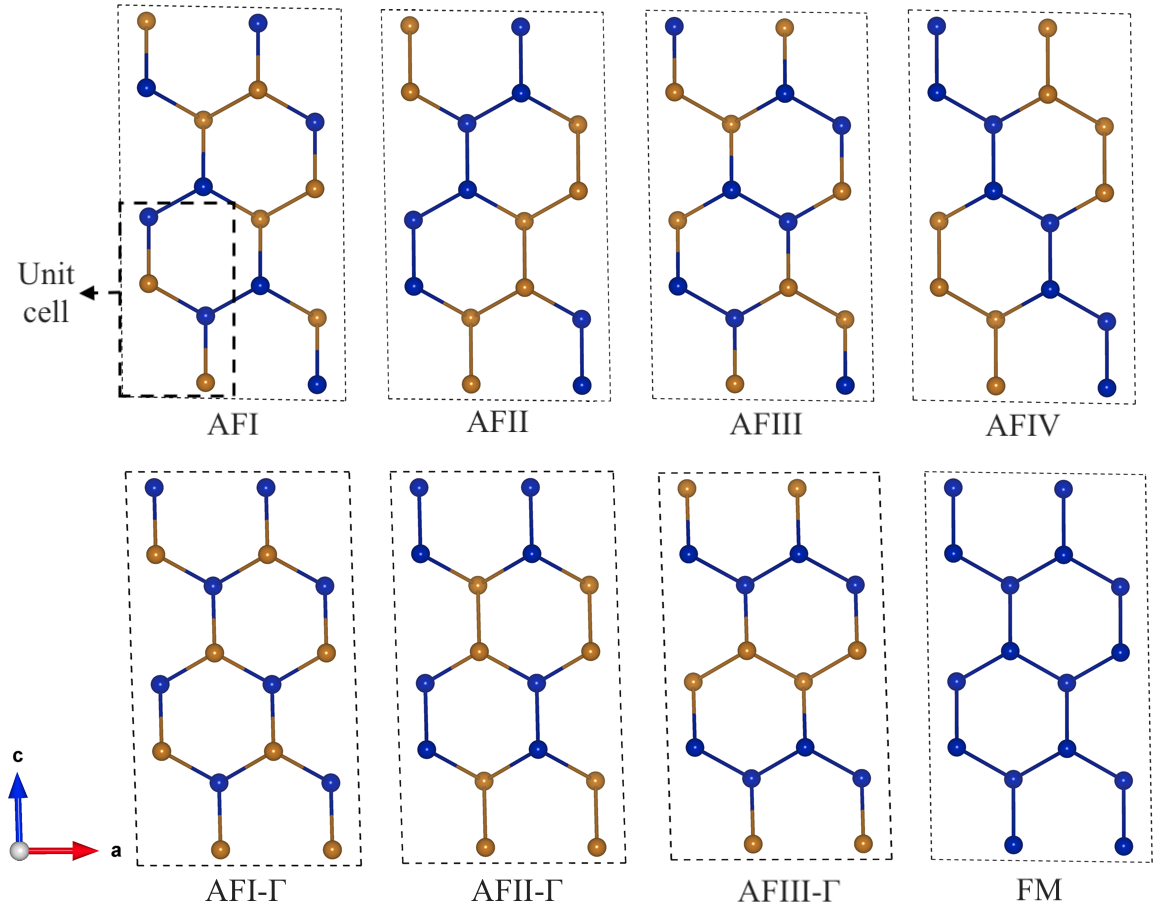}
\centering
\caption[magUnitcell]{\label{fig:SMfigure1} The eight magnetic structures considered shown in the $2\times1\times2$ cell. For clarity, only the Co atoms are shown, color blue for spin up and brown spin down polarization.  A description of the 4 Co atoms in the unit cell is shown. } 
\end{figure}

\begin{table} [!h]
\begin{threeparttable}		
\caption{\label{tab:SMtable1} Four maximal magnetic space groups for the parent space group P2$_1$/c,  and the magnetic structure of the 4 symmetry equivalent magnetic Co atoms that describes one quarter of the magnetic unit cell.  } 
		\begin{ruledtabular}
			\begin{tabular}{ c  @{\hspace{0.8em}} l @{\hspace{0.8em}} l @{\hspace{0.8em}} l @{\hspace{1em}}l  } 
				& \multicolumn{4}{c}{Magnetic structure~\tnotex{tnxx59}} \\
				\cline{2-5} \\
				
				Label~\tnotex{tnxx58} & Co1 & Co2& Co3 &  Co4 \\ 
				\colrule
				AFI     & ($m_x$, $m_y$, $m_z$)     & ($-m_x$, $m_y$, $-m_z$)    & ($-m_x$, $-m_y$, $-m_z$)  & ($m_x$, $-m_y$, $m_z$) \\
				AFII    & ($m_x$, $m_y$, $m_z$)     &($m_x$, $-m_y$, $m_z$)      &  ($-m_x$, $-m_y$,$-m_z$)  & ($-m_x$, $m_y$, $-m_z$) \\
				AFIII   & ($m_x$, $m_y$, $m_z$)     &($-m_x$, $m_y$, $-m_z$)     &  ($m_x$, $m_y$, $m_z$)    & ($-m_x$, $m_y$, $-m_z$) \\
			    AFIV    & ($m_x$, $m_y$, $m_z$)     & ($m_x$, $-m_y$, $m_z$)     & ($m_x$, $m_y$, $m_z$)      & ($m_x$,$-m_y$, $m_z$) \\
			\end{tabular} 
		\end{ruledtabular}
		\begin{tablenotes}
		    \item[a]  \label{tnxx59}Co1, Co2, Co3, Co4 at: (x, y, z), (-x+3/4, y+1/2, -z+1/4), (-x+1/2, -y, -z+1/2) and (x+3/4, -y+1/2, z+1/4) positions respectively in one quarter of the magnetic supercell.
			\item[b]  \label{tnxx58} Label considering collinear calculations with magnetization along the \textbf{z} axis ($m_x = m_y=0$).
		\end{tablenotes} 
	\end{threeparttable}
	\
\end{table}
In addition, we have considered four other magnetic structures that result from the propagation at $\Gamma=$(0, 0, 0) of the unit cells for the AF structures discussed above. These new four structures are labelled AFI-$\Gamma$, AFII-$\Gamma$, AFIII-$\Gamma$ and FM (i.e AFIV-$\Gamma$).  All the magnetic structures in the 16 Co cell considered are shown along the $a-c$ plane in Fig~\ref{fig:SMfigure1}, while their relative DFT energies per formula unit for relaxed atomic positions are reported in Table~\ref{tab:SMtable2}. The lowest energy stable structure is the AFI followed by AFII,  AFIII and AFIV. However, if we consider fixed atomic positions at the AFI relaxed positions, the total DFT energy order of the AFII and AFIII structures are reversed . Further results shown are for the AFI structure.

\begin{table} [!h]
\begin{threeparttable}		
\caption{\label{tab:SMtable2}  Total energy differences for the magnetic structures with respect to AFI are presented for both DFT and DFT+U optimized atomic positions in meV/f.u.  } 
		\begin{ruledtabular}
			\begin{tabular}{ c  @{\hspace{1em}} l  @{\hspace{1em}}  l } 
				Label & $\Delta{E}_{DFT}$~\tnotex{tnxx60}  &  $\Delta{E}_{DFT+U}$~\tnotex{tnxx60} \\ 
				\colrule
				AFI    & 0.0 & 0.0 \\
				AFII   & 2.164 & 0.208 \\
				AFIII  & 4.744 &  1.640   \\
			    AFIV   & 8.365 & 3.830   \\
			    AFI-$\Gamma$    & 6.085& 5.885 \\
				AFII-$\Gamma$   & 7.904 & 5.051\\
				AFIII-$\Gamma$  & 11.362 &  7.507   \\
			    FM   & 44.427 & 32.169  \\
			\end{tabular} 
		\end{ruledtabular}
		\begin{tablenotes}
			\item[a]  \label{tnxx60} Reported energy differences in meV per formula unit.
		\end{tablenotes} 
	\end{threeparttable}
	\
\end{table}

Results of  the structural optimization with both DFT and DFT+U are presented in Table~\ref{tab:SMtable3} for the AFI structure. The calculated lattice parameters and bond distances are in good agreement with experiment~\cite{jin2006}. The  optimization with fixed lattice parameter (column 4 of Table~\ref{tab:SMtable3})  gives Co$-$O and Co$\dotsb$Co distances in most agreement with experiment and was adopted for further calculations.

\begin{table} [!h]
 \begin{threeparttable}
\caption{\label{tab:SMtable3} Optimized lattice parameters and atomic distances from DFT, DFT+U and experiment are presented. Oxygen atom labels are defined here to be used also in Tab~\ref{tab:SMtable5}.}
\begin{ruledtabular}
 \begin{tabular}{ l @{\hspace{0.8em}} l @{\hspace{0.8em}} l @{\hspace{0.8em}} l @{\hspace{0.8em}} l@{\hspace{0.8em}} l} 
                              & DFT                          &DFT+U         & DFT+U & Exp.~\cite{jin2006} \\ 
     \colrule
    a (\AA)              &  6.103                         & 6.105         &  ~\tnotex{tnxx62}   &  5.9176\\
    b (\AA)              &  7.686                         & 7.694         & ~\tnotex{tnxx62}    &  7.6192\\
    c (\AA)              &  10.503                       & 10.522        &  ~\tnotex{tnxx62}   &  10.3970\\
    $\beta$ (\si{\degree})  &  93.519                       & 93.412        &  ~\tnotex{tnxx62}   &  91.734\\
   $d_{Co-O_1}$ (\AA)     &  1.793                         & 1.799          & 1.795                     & 1.775  \\
  $d_{Co-O_2}$  (\AA)      &  1.809                         & 1.817          & 1.798                     &  1.778  \\
   $d_{Co-O_3}$ (\AA)      &  1.825                         & 1.834          &  1.816                    &  1.794  \\
  $d_{Co-O_4}$  (\AA)      &  1.835                         & 1.842          &  1.832                    &  1.813  \\
\hline    
    $d_{Co \dotsb Co}$ (\AA) &  4.787,  & 4.789,         &  4.756,                   &  4.765,\\
                              &  5.023,                       & 5.031,         &  4.865,                   &  4.871,\\
                             & 5.216,                        & 5.217,         &  5.192,                   &  5.193,\\
                             &  5.407,                        & 5.415,        &  5.364,                   &  5.357, \\
                             &  5.618,                        & 5.630,        &  5.461,                   &  5.452,\\
                             & 5.924,                        &5.938,          &  5.917,                   &  5.916,\\
                             &  6.103,                       & 6.105,         &  5.918,                   &  5.918,\\
                             & 6.247,                        & 6.251,         &  6.072,                   &  6.072,\\   
    \end{tabular} 
  \end{ruledtabular}
    \begin{tablenotes}
             \item[a] \label{tnxx62} Fixed to experimental value~\cite{jin2006}.    
 \end{tablenotes} 
 \end{threeparttable}
  \
\end{table}

\section{\label{sec:appendix3} Search for occupation of Co spin states}
DFT+U suffers from multiple local minima that have been shown to prevent the individuation of the true ground state in many compounds.
In order to fully explore the possibility of unconventional spin states in \bco, we explored several initial occupation possibilities for Co 3$d$, corresponding to high , intermediate and low spin.
In all simulations, the self consistent procedure converged to the occupancies reported in Tab.~\ref{tab:SMtable4}, \ref{tab:SMtable5} and in the main text.
 
\begin{table}[!h]
 \begin{threeparttable}
\caption{\label{tab:SMtable4} Occupancy of the Co$-d$ states (distorted tetrahedral) with DFT+U.}
\begin{ruledtabular}
 \begin{tabular}{ l @{\hspace{0.8em}} l @{\hspace{0.8em}} l @{\hspace{0.8em}} l @{\hspace{0.8em}} l@{\hspace{0.8em}} l@{\hspace{0.8em}} l@{\hspace{0.8em}} l} 
                              & d$_{x^2-y^2}$      &  d$_{z^2}$      &  d$_{xz}$   &  d$_{yz}$     & d$_{xy}$    &  total\\ 
     \colrule
    Majority spin     &       0.998  &   0.998 &  0.994 &  0.994 & 0.992  & 4.97649\\
     Minority spin    &       0.608  &   0.600 & 0.569  & 0.360 & 0.330  & 2.46654\\   
     total   &      &             &             &           &            &7.44303
       \end{tabular} 
  \end{ruledtabular}
 \end{threeparttable}
  \
\end{table}

\begin{table}[!h]
 \begin{threeparttable}
\caption{\label{tab:SMtable5}Occupancy of the O$-p$ states  (distorted tetrahedral) with DFT+U.}
\begin{ruledtabular}
 \begin{tabular}{l @{\hspace{0.8em}} l @{\hspace{0.8em}} l @{\hspace{0.8em}} l @{\hspace{0.8em}} l @{\hspace{0.8em}} l@{\hspace{0.8em}} l} 
                    &         & p$_x$      &  p$_z$      &  p$_y$   &  total\\ 
     \colrule
    O$_1$ &Majority spin      &  0.967     &    0.970  &    0.962  &  2.89919\\
                &Minority spin     &      0.851  & 0.835    & 0.801   & 2.48588\\   
                &  total                  &      &             &                      &5.38507\\
                \\
                
     O$_2$ & Majority spin     &    0.961     &  0.966  &     0.969  & 2.89517\\
              & Minority spin     &      0.788   & 0.851   & 0.865     & 2.50295\\               
                & total                  &      &             &                      &5.39812\\
                \\
     O$_3$&Majority spin          & 0.961  &     0.966  &    0.964  & 2.88981\\
                & Minority spin     & 0.787   &    0.870    &   0.866    &  2.52242\\    
               &  total                 &      &             &                      &5.41223\\
               \\
       O$_4$ &Majority spin       &  0.963       &    0.962  &     0.966 & 2.89072\\
                  & Minority spin     &    0.875       &   0.789  &     0.865 &2.52890\\    
                  & total                   &      &             &                      &5.41961
       \end{tabular} 
  \end{ruledtabular}
 \end{threeparttable}
  \
\end{table}

\begin{figure}[!h]
\centering
\includegraphics[width=17.0cm]{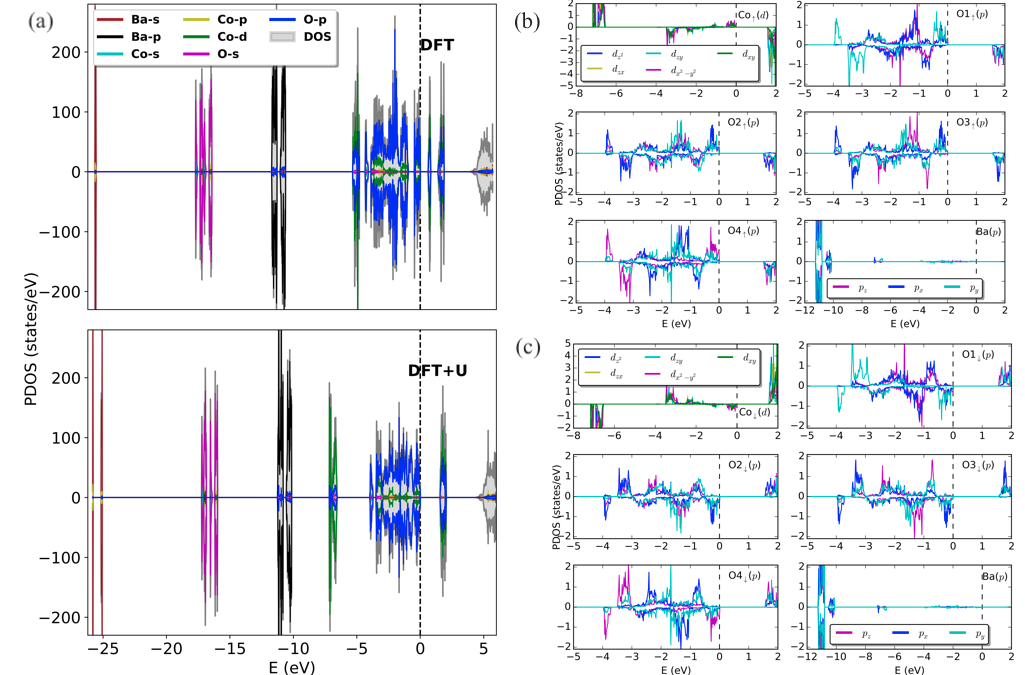}
\phantomsubfloat{\label{fig:SMfigure2a}}
 \phantomsubfloat{\label{fig:SMfigure2b}}
 \phantomsubfloat{\label{fig:SMfigure2c}}
\vspace{-2\baselineskip}%
\centering
\caption[Unitcell ]{\label{fig:SMfigure2}(a) DOS and PDOS spanning all available energy ranges.  Representative DFT+U PDOS for $d$ and $p$ orbitals of a CoO$_4$ with (b)majority spin channel (up) and Ba.  (c) minority spin channel(down)  and Ba. The valence band maximum is shifted to 0 eV. } 
\end{figure}

\section{\label{sec:appendix2pd}Projected density of States }
In Fig~\ref{fig:SMfigure2}, we plot the density of states (DOS) spanning over all the available energy levels and the projection  of the DOS to the atomic orbitals, showing the contribution of each atom.

\section{\label{sec:appendixml}Details of Wannier functions and exchange coupling.}
\begin{figure}[!h]
\centering
\includegraphics[width=18.0cm]{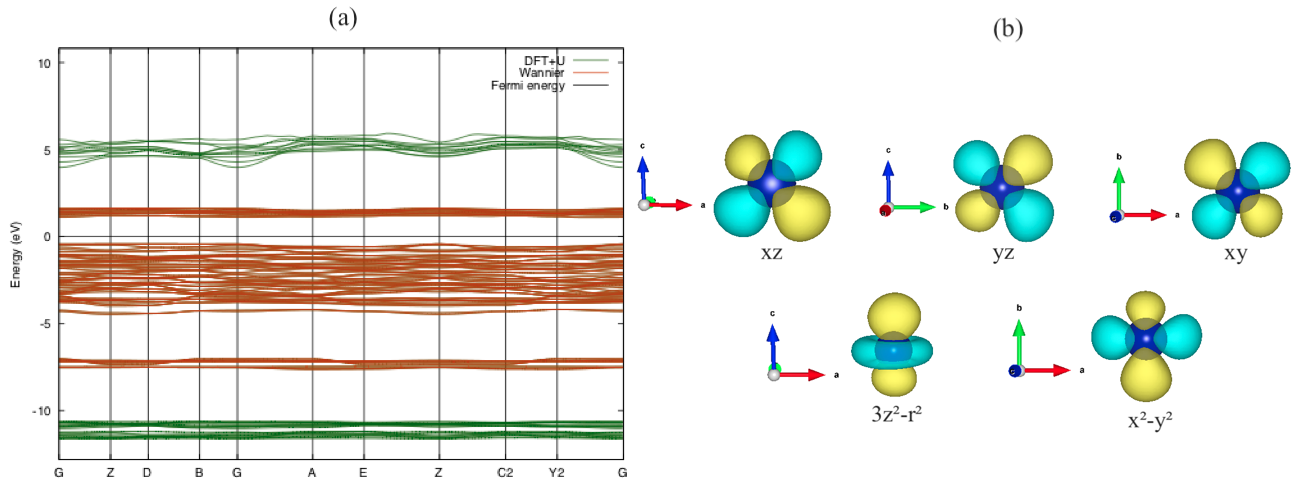} \\
\phantomsubfloat{\label{fig:SMfigure3a}}
 \phantomsubfloat{\label{fig:SMfigure3b}}
\caption[band plots]{\label{fig:SMfigure3} (a) DFT+U bands and the interpolated Wannier bands (spin up) in the conventional unit cell. (b) Isosurface plot of the maximally localized Wannier functions localized and centred on Co and labelled by their $d$ orbital character using the magnetic unit cell. This plot was reproduced with Vesta~\cite{momma2011}.}
\end{figure}

\begin{figure}[!h]
\centering
\includegraphics[width=8.5cm]{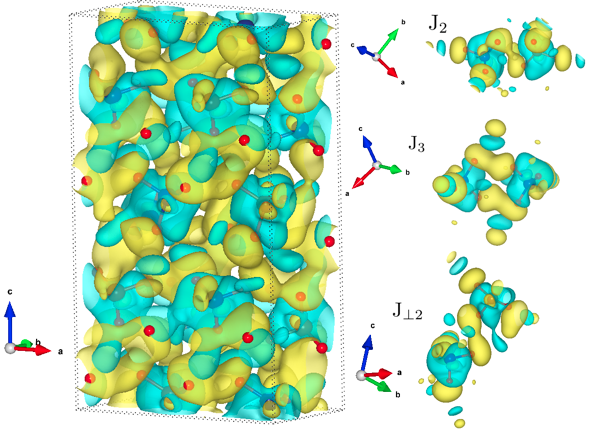}
\caption[band plots]{\label{fig:SMfigure4} Plot of the Wannier function (spin-up) showing the overlap of the $p$ orbitals  for the magnetic unit cell  and representative plots of the Wannier on adjacent tetrahedra complex  for  J$_2$, J$_3$ and (d) J$_{\bot 2}$ interactions.}
\end{figure}

In order to quantify from first principles the exchange coupling interactions and  discuss the exchange mechanism, first, we obtain a local basis by mapping the spin polarized bands to  Wannier functions (WF)  using the WANNIER90 code~\cite{mostofl2008,pizzi2020}. Following careful tests of various combination of different orbitals, the WF were generated by the projection of the Co 3$d$ and O 2$p$ orbitals on the subspace of 272 Bloch bands.  We ensured to obtain  localized atomic orbitals and matching Wannier and DFT bands in the subspace considered.  Fig.~\ref{fig:SMfigure3a} shows the bands while Fig.~\ref{fig:SMfigure3b} shows the plot of the WF  localized and centered on the Co,  showing the localized orbital character. In Fig.~\ref{fig:SMfigure4}, the Wannier functions are plotted in the magnetic unit cell and shown also are the representatives of the exchange coupling indicating the exchange mechanism path and overlap of WF of the mediating O atoms from adjacent tetrahedra complex.

In order to describe the magnetic interactions of the system we adopt the widely used Heisenberg model~\cite{heisenberg1928}. The Heisenberg Hamiltonian is written as; 
\begin{equation}
H = - \sum_{i\neq j}{ J_{ij} \mathbf{S}_i  \cdot  \mathbf{S}_j}.\label{eq:Heisen}
\end{equation}
where J$_{ij}$  is the  inter atomic \emph{isotropic} exchange coupling constants) and  $\mathbf{S}_i$   and $\mathbf{S}_j$ the spin vectors normalized to 1 at sites i and j.
The isotropic exchange coupling constant is computed using the Green's function approach~\cite{liechtenstein1987,mazurenko2007,korotin2015} which has been used successfully to describe the exchange constants and magnetic interactions in  a number of oxides~\cite{mazurenko2005,mazurenko2007,pchelkina2014,korotin2015,logemann2017,xu2021}.

Following the approach in Ref.\cite{korotin2015,xu2021} for plane-wave DFT pseudopotential calculations and using the implementation in the TB2J code~\cite{xu2021}, the isotropic contribution to the exchange constants, $J_{ij}$ were calculated by the Green's function method with the Wannier functions as localized basis (that allows to map the calculations with DFT+U to the Heisenberg model). In the TB2J code, the spin operators in Eq.~\ref{eq:Heisen}  are  replaced  with unit vectors (or normalized to 1)  that has directions as those of the magnetic moment on sites i and j. Thus, for comparison with the classical Heisenberg Hamiltonian whose spins are not normalized to 1, the reported exchange coupling constants (from the TB2J code) have  to be renormalized.

\begin{table*}
\begin{threeparttable}		
\caption{\label{tab:SMtable6}  Neighbour inter-tetrahedra  exchange coupling contributions including the interactions with the ligand spin, calculated using the Green's function approach with the ligand spins also treated as localized. } 
		\begin{ruledtabular}
			\begin{tabular}{ l @{\hspace{1em}}  p{2.0cm}   p{2.0cm}  p{2.0cm}   p{2.0cm}  p{2.0cm}  p{2.0cm}   p{2.0cm}   p{2.0cm}  p{2.0cm}  } 			
				Label~\tnotex{tnxx85} & J$_1$ (meV) \newline distance \AA & J$_2$ (meV) \newline  distance \AA & J$_3$ (meV) \newline  distance \AA & J$_4$ (meV) \newline  distance \AA &  J$_5$ (meV) \newline  distance \AA &  J$_{\bot 1}$ (meV) \newline  distance \AA & J$_{\bot 2}$ (meV)  \newline  distance \AA & J$_{\bot 3}$ (meV)  \newline distance \AA\\ 
				\colrule
				Co$\dotsb$Co &-0.241 \newline  4.756  & -0.337  \newline  4.865   & -1.902  \newline  5.364 & -2.723  \newline 5.461  & -2.334  \newline 5.918 &  -1.016 \newline 5.192 & -0.757 \newline 5.917 & -0.719  \newline 6.072 \\
				\\
				O$_1\dotsb$O$_1$ & 0.0100 \newline 3.577~\tnotex{tnxx81} & 0.0225  \newline 3.245~\tnotex{tnxx81} & -0.0041 \newline 8.573  & -0.0043 \newline 8.440 & -0.0003 \newline 5.917 & 0.0008  \newline 5.192 & -0.0004 \newline 7.122  & -0.0022 \newline 7.251  \\
				O$_1\dotsb$O$_2$ & 0.0024  \newline 3.747 & 0.0008 \newline 5.510 & -0.0068 \newline 6.257  & -0.0022 \newline 7.447  & -0.0069 \newline 5.024 & 0.0002 \newline 6.774  & -0.0002 \newline 6.322 & 0.0046 \newline 4.822 \\
				O$_1\dotsb$O$_3$ & 0.0023 \newline 5.615 &  0.0091 \newline 3.609  & -0.0008 \newline 7.272  & -0.0109 \newline 5.864  &-0.0059 \newline 7.757  & 0.0001 \newline 7.226 & 0.0154 \newline 5.002 & -0.0020 \newline 6.567 \\
				O$_1\dotsb$O$_4$ & 0.0085 \newline 3.652 & 0.0054 \newline 3.882   & -0.0059 \newline 5.991 & -0.0043 \newline 6.134  & -0.0001 \newline 6.343  & -0.0003 \newline 8.175  & -0.0029 \newline 7.706 & -0.0017 \newline 7.566  \\
				O$_2\dotsb$O$_1$ & 0.0024  \newline 3.748  & 0.0008 \newline 5.510  & -0.0068 \newline 6.257  & -0.0022 \newline 7.448 & -0.0023 \newline 7.899  & 0.0587 \newline 3.208~\tnotex{tnxx81} & -0.0013 \newline 6.468 & -0.0015 \newline 7.891  \\
				O$_2\dotsb$O$_2$ &-0.0002  \newline 5.735  & -0.0023 \newline 8.233  & -0.0026 \newline 4.733  & -0.0005 \newline 7.569  & -0.0093 \newline 5.918 & 0.0006 \newline 4.998  & 0.0025 \newline 5.943  & -0.0011 \newline 6.098  \\
				O$_2\dotsb$O$_3$ & 0.0011  \newline 6.456  & 0.0034 \newline 6.462  & -0.0009 \newline 5.238  & -0.0006 \newline 5.246  & -0.0077 \newline 8.868 & 0.0037 \newline 4.508 & -0.0018 \newline 5.601  & -0.0013 \newline 8.249  \\
				O$_2\dotsb$O$_4$ & 0.0182 \newline 4.918 & 0.0009 \newline 6.671  & 0.0094 \newline 3.339~\tnotex{tnxx81} & -0.0015 \newline 5.609 & -0.0037 \newline 7.616 & 0.0005 \newline 6.031  & -0.0009 \newline 8.018  & -0.0010 \newline 8.985 \\
				 O$_3\dotsb$O$_1$ &0.0023 \newline 5.615 & 0.0091 \newline 3.609  & -0.0008 \newline 7.272  &  -0.0109 \newline 5.863 & -0.0068 \newline 5.263  & 0.0149 \newline 3.587 & -0.0002 \newline 7.288  & -0.0000 \newline 6.224 \\
				O$_3\dotsb$O$_2$ & 0.0011 \newline 6.456 & 0.0034 \newline 6.462 & -0.0009 \newline 5.238  & -0.0006 \newline 5.245  &  0.0725 \newline 3.004~\tnotex{tnxx81}  & 0.0012 \newline 4.071 & -0.0005 \newline 7.589  & 0.0010 \newline 4.963 \\
				O$_3\dotsb$O$_3$ & -0.0016  \newline 8.252 & -0.0022 \newline 5.776 & -0.0008 \newline 7.079 & -0.0302  \newline 3.921 & -0.0098 \newline 5.918  & 0.0009 \newline 5.414  & 0.0029 \newline 6.020  &  -0.0002 \newline 6.173 \\
				O$_3\dotsb$O$_4$ & 0.0045 \newline 6.481  & 0.0113 \newline 5.245 & 0.0015 \newline 5.023  &  0.0266 \newline 3.273~\tnotex{tnxx81} & 0.0016 \newline 4.794 & -0.0014 \newline 6.069 & -0.0009 \newline 8.814 & 0.0007 \newline 7.702  \\
				O$_4\dotsb$O$_1$ &0.0085 \newline 3.652  & 0.0054 \newline 3.882 & -0.0059 \newline 5.991 & -0.0043 \newline 6.134  & -0.0022 \newline 6.945 & -0.0019 \newline 5.356  & -0.0011 \newline 4.539  & -0.0011 \newline 5.143 \\
				O$_4\dotsb$O$_2$ & 0.0182  \newline 4.918 & 0.0009 \newline 6.671  & 0.0094 \newline 3.339~\tnotex{tnxx81} & -0.0015 \newline 5.609 & -0.0080 \newline 5.371  &  0.0014 \newline 5.816 & -0.0014 \newline 4.862 & 0.0173 \newline 3.304~\tnotex{tnxx81} \\
				O$_4\dotsb$O$_3$ & 0.0045 \newline 6.481  & 0.0113 \newline 5.245  & 0.0015 \newline 5.022 & 0.0265 \newline 3.273~\tnotex{tnxx81}  & -0.0039 \newline 7.966 &-0.0007 \newline 6.167  & 0.0246 \newline 3.289~\tnotex{tnxx81} & -0.0011 \newline 5.737  \\
				O$_4\dotsb$O$_4$ & 0.0014 \newline 5.684 & -0.0005 \newline 6.168 &  -0.0032 \newline 4.089 & -0.0238 \newline 4.738&0.0055 \newline 5.918  & -0.0008 \newline 7.869&-0.0007 \newline 5.991  & 0.0011 \newline  6.145  \\
				Co$\dotsb$O$_1$ & 0.0336 \newline 3.806 & 0.0659 \newline 3.725  & -0.1108 \newline 6.922 & -0.1255 \newline 6.878  &  -0.0765 \newline 6.316  & 0.0122 \newline 4.063 & -0.0234 \newline 6.230 & -0.0257 \newline 6.506 \\
				Co$\dotsb$O$_2$ & 0.0094  \newline 4.952 & -0.0156 \newline 6.518 & -0.0654 \newline 4.728 & -0.0460 \newline 6.350  & -0.1166 \newline 4.619 & -0.0089 \newline 5.243 & -0.0145 \newline 6.000 & -0.0294 \newline 4.576 \\
				Co$\dotsb$O$_3$ & -0.0109 \newline 6.486 & 0.0425 \newline 5.022 &-0.0178 \newline 6.013 & -0.1681 \newline 4.393  & -0.1326 \newline 7.497 & -0.0090 \newline 5.668  & 0.0347 \newline 4.770 & -0.0170 \newline 6.520 \\
				Co$\dotsb$O$_4$ & 0.0588  \newline 4.910 & 0.0157 \newline 5.244 & -0.0794 \newline 4.404 & -0.1010 \newline 4.773 & -0.0525 \newline  6.003 & -0.0418 \newline 6.895  & -0.0244 \newline 7.504   & -0.0309 \newline 7.472  \\
				O$_1\dotsb$Co & 0.0336  \newline 3.806  & 0.0659 \newline 3.725 & -0.1108 \newline 6.922 & -0.1254 \newline 6.878 & -0.0994 \newline 6.048  & -0.0114 \newline 6.617  & -0.0373 \newline 6.368  & -0.0323 \newline 6.384 \\
				O$_2\dotsb$Co & 0.0094  \newline 4.952 & -0.0156 \newline 6.519  & -0.0654 \newline 4.728  & -0.0462 \newline 6.350  & -0.1220 \newline 7.428 & 0.0178  \newline 4.379 & -0.0249\newline 6.294 & -0.0282 \newline 7.641  \\
				O$_3\dotsb$Co & -0.0109  \newline 6.486  & 0.0425 \newline 5.022 & -0.0178 \newline 6.012 & -0.1683 \newline 4.393 & -0.0977 \newline 4.519 & -0.0312 \newline 4.478  & -0.0211 \newline 7.251  & 0.0140 \newline 6.046  \\
				O$_4\dotsb$Co & 0.0588 \newline 4.910  & 0.0157 \newline 5.244 & -0.0794 \newline 4.404  & -0.1011 \newline 4.773 & -0.0727 \newline 6.381  &-0.0211 \newline 6.056 & -0.0313 \newline 4.390 & 0.0002 \newline 4.845 \\
			\end{tabular} 
		\end{ruledtabular}
		\begin{tablenotes}
		    \item[a]  \label{tnxx85}Site interaction
		     \item[b]  \label{tnxx81}Distance of inter-tetrahedra mediating O$\dotsb$O for each exchange coupling parameter.
		\end{tablenotes} 
	\end{threeparttable}
	\
\end{table*}

\section{\label{sec:appendix6} Oxygen spin contribution to the magnetic interaction}
\begin{figure}[!h]
\centering
\includegraphics[width=6cm]{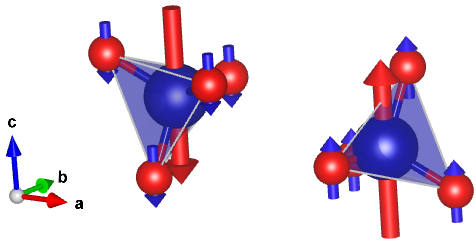}
\caption[band plots]{\label{fig:SMfigure5} Illustration of two neighbour tetrahedra complex including the spin on oxygen sites parallel to the centre Co.}
\end{figure} 
In this section we tabulate the exchange coupling constants calculated using the Green's function while treating the induced spin at the oxygen sites as localized.  An illustration of two neighbour tetrahedra complex with the induced spin on the ligand sites is shown in Fig.~\ref{fig:SMfigure5}. In order to consider the ligand spin contribution to the magnetic interaction, first the exchange constants J in Table~\ref{tab:SMtable6} are listed considering the distance between Co$\dotsb$Co  pairs. For each Co$\dotsb$Co pair, we consider the ligand contributions, which are all the inter-tetrahedral site interactions and include; the Co$\dotsb$Co, 4 Co$\dotsb$O, 4 O$\dotsb$Co and 16 O$\dotsb$O interactions. The DFT+U calculated exchange coupling constants and their respective distances are listed in Table~\ref{tab:SMtable6}. From, Table~\ref{tab:SMtable6} the coupling constants show that the  Co$\dotsb$Co couplings are the dominant interactions in the system, while the Co$\dotsb$O, O$\dotsb$Co and O$\dotsb$O parameters are order(s) of magnitude smaller relative to those of the corresponding  Co$\dotsb$Co.  For instance, for J$_5$, the \emph{magnitude} of the strongest ligand interactions for
Co$\dotsb$O/O$\dotsb$Co and O$\dotsb$O are -0.122 and 0.073 meV respectively, which are small relative to -2.334 meV of Co$\dotsb$Co. This order of magnitude is similar for the other couplings. Thus indicating that the Co$\dotsb$Co couplings dominate the exchange interaction. It is also important to note that most of the O$\dotsb$O couplings are vanishing except the short distanced ones (listed in Table~3 (main text)), which are  all coupled ferromagnetically and they also mediate the indirect exchange antiferromagnetic interactions between two neighbour tetrahedra complexes.  Such that the absence of Wannier functions of any of the O$\dotsb$O pairs does not result in overlap between the tetrahedra complexes and hence vanishing exchange interactions. 

Also, we consider all the Co$-$O and O$-$O interactions within each tetrahedron complex (intra-tetrahedron interaction). The DFT+U calculated values are listed in Table~\ref{tab:SMtable7}.  The Co$-$O and O$-$O interactions are all ferromagnetic, showing strong coupling.

We further invoked a more refined downfolding procedure~\cite{solovyev2021} to transfer the effects of the ligand spins to those of the Co$\dotsb$Co interactions. These results are summarized and discussed in Table 3 (main text) and corresponding sections in the main text.

\begin{table} 
\begin{threeparttable}		
\caption{\label{tab:SMtable7}  Intra-tetrahedron  exchange coupling contributions. } 
		\begin{ruledtabular}
			\begin{tabular}{ l @{\hspace{1em}}  p{1.5cm}   p{1.5cm}  } 			
				Label~\tnotex{tnxx80} & distance \AA &  J (meV)  \\ 
				\colrule
				Co$-$O$_1$ &1.795   & 59.39  \\
				Co$-$O$_2$ &1.798   & 56.17  \\
				Co$-$O$_3$ &1.816   & 49.86\\
				Co$-$O$_4$ &1.832 & 48.68  \\
				O$_1-$O$_2$ &2.966   & 1.61  \\
				O$_1-$O$_3$ &2.987   & 1.44  \\
				O$_1-$O$_4$ &3.036   & 1.58  \\
				O$_2-$O$_3$ &2.969  & 1.46  \\
				O$_2-$O$_4$ &2.900 & 1.38  \\
				O$_3-$O$_4$ &2.863   & 1.29 
			\end{tabular} 
		\end{ruledtabular}
		\begin{tablenotes}
		    \item[a]  \label{tnxx80}Site interaction
		\end{tablenotes} 
	\end{threeparttable}
	\
\end{table}

\section{\label{sec:appendix5} Magnetic Structure Stability with the Heisenberg model}
In other to investigate the stability of the considered magnetic structures within the Heisenberg model, we calculated the energy ($\mathcal{E}$) per spin in the unit cell as a function of the propagation vector $\mathbf{q}$ and of the spin structure, given by: 
\begin{equation}
\mathcal{E}(\mathbf{q})=- \frac{1}{2N} \sum_{i j} {J_{i j} (\mathbf{q}) \mathbf{S}_{\mathbf{q} i}  \cdot  \mathbf{S}_{-\mathbf{q} j}}, \label{eq:energq}
\end{equation}

where N is the number of i, j sites and $J(\mathbf{q})$, the Fourier transform of the exchange coupling parameters $J(\mathbf{R})$ ($\mathbf{R}$ is the lattice sites)  between spins $\mathbf{S}$ is given as: 

\begin{equation}
J(\mathbf{q}) = \sum_{\mathbf{R}} {J(\mathbf{R}) e^{i \mathbf{q} \cdot \mathbf{R}}}
\end{equation}

and 

\begin{equation}
\mathbf{S}_{\mathbf{q}} =\frac{1}{N} \sum_\mathbf{R}{\mathbf{S}_{\mathbf{R}} e^{i \mathbf{q} \cdot \mathbf{R}} }
\end{equation}
with the spin spiral in  $\mathbf{q}$ about a rotation axis in the $z$ direction given as;
\begin{equation}
\mathbf{S}_{\mathbf{R}} = S(\cos(\mathbf{q} \cdot \mathbf{R}), \sin(\mathbf{q} \cdot \mathbf{R}), 0)
\end{equation}
 S has the magnitude of 1 in our case and we ensured same spin length for all the lattice sites.
 
\begin{figure}[!h]
\centering
\includegraphics[width=16.0cm]{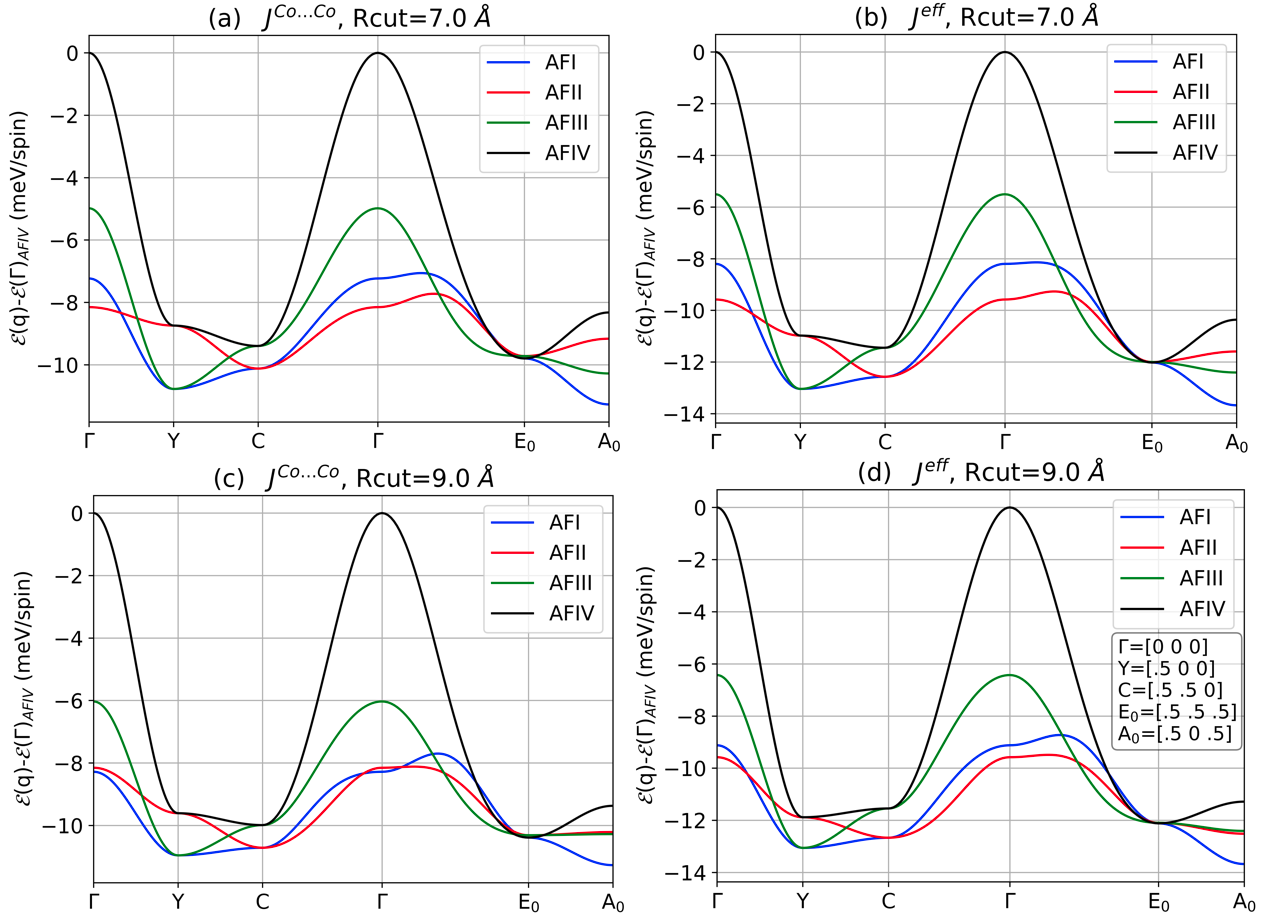}
\phantomsubfloat{\label{fig:SMfigure6a}}
 \phantomsubfloat{\label{fig:SMfigure6b}}
 \phantomsubfloat{\label{fig:SMfigure6c}}
 \phantomsubfloat{\label{fig:SMfigure6d}}
\vspace{-2\baselineskip}%
\centering
\caption[SMfigure6 ]{\label{fig:SMfigure6} Figures a-d show the Heisenberg energy (in meV/spin) with respect to $\mathbf{q}$ along selected path and the four AF structures considered in this work. We used the exchange coupling constants (with cutt-off radius (Rcut) upto 7~\AA{} and 9~\AA) computed by the two DFT+U approaches:  $J^{\mathrm{Co} \dotsb \mathrm{Co}}$  and  $J^{\mathrm {eff}}$, excluding and including the ligand spin contributions respectively. The energy axis (in meV/spin) is shifted relative to $\mathcal{E}(\Gamma)_{AFIV}$ (which is also the energy of the ferromagnetic structure). } 
\end{figure}

Fig.~\ref{fig:SMfigure6} shows the plot of the Heisenberg energy (Eq.~\ref{eq:energq}) with respect to $\mathbf{q}$ and the four AF structures considered in this work, using the computed exchange coupling constants ($J^{\mathrm{Co} \dotsb \mathrm{Co}}$  and  $J^{\mathrm {eff}}$) described in Table 1 of the main text.   

In the main text we have considered only the stronger interactions, within a cutoff sphere radius (Rcut) of 7~\AA. In fig.~\ref{fig:SMfigure6}  we compare Eq.~\ref{eq:energq} for the same cut-off (panel a and b) and for a larger cutoff of 9~\AA{} (panel c and d) which include farther very weak interactions. The figure shows that in all cases the lowest energy state is the AFI structure at the antiferromagnetic wave vector (0.5, 0.0, 0.5) (the A$_0$ point in fig.~\ref{fig:SMfigure6}). This shows that classically, within the spin configurations considered the lowest energy state is indeed antiferromagnetic in agreement with experiment and the AFI structure is the lowest energy structure in agreement with collinear DFT+U calculations. The Hamiltonian matrix constructed with the AFI structures are real and positive-defined at the (0.5, 0.0, 0.5) propagation vector.

 We notice, in addition, that at $\mathbf{q} =$ (0.5,0.0,0.5), AFIII has lower energy than the AFII in the Heisenberg model with Rcut = 7~\AA\ (Fig.~\ref{fig:SMfigure6a} and \ref{fig:SMfigure6b}), in contrast with the DFT+U (DFT) results from relaxed structures.
 As a matter of facts DFT calculations at fixed AFI atomic positions give the same energy order as the Heisenberg model. We further notice that at Rcut = 9~\AA\ (Fig.~\ref{fig:SMfigure6c} and \ref{fig:SMfigure6d}), AFII and AFIII energies become degenerate at (0.5,0.0,0.5). Also, all the AF$-\Gamma$ structures have higher energy to their (0.5,0.0,0.5) counterparts.

\section{\label{sec:appendix4}Mean-Field estimation of T$_N$ and $\Theta$ }

 Mean-field estimate of T$_N^m$ and $\Theta^m$ using the reported calculated isotropic contribution to the exchange constants J  were performed with the following relation~\cite{solyom2007};
\[T^m_N= \frac{1}{3K_B} \left( \sum_{i=\{a\}}\eta_{i}|J_{i}^{*}|+ \sum_{i=\{b\}}\eta_{i}J_{i}^{*} \right) \]  and \[\Theta^m= \frac{1}{3K_B} \sum_{i=\{a,b\}}\eta_{i}J_{i}^{*}\]

where \{a\} =\{2, 4, 5, $\bot 1$, $\bot 2$\}  represents index for  interaction between anti-parallel spins on Co, while  \{b\} =\{1, 3, $\bot 3$\}  represents  interaction between parallel spins. $J^*$=J$\cdot$S(S+1)/$S^2$ allows to scale the interactions from classical to quantum model using the value of the spin, S. $K_B$ is the Boltzmann constant, $\eta$ is the number of nearest neighbour interaction for each $J$. The calculated parameters for different spins (S) are listed in table~\ref{tab:SMtable8}

\begin{table}[!h]
    \centering
    \caption{Calculated $T_N^m$ and $|\Theta^m|$ assuming an ionic, Cobalt only picture S=3/2 and the hybridized Cobalt-ligand picture, S=5/2 using $J^{\mathrm{Co} \dotsb \mathrm{Co}}$ and $J^{\mathrm {eff}}$ reported in Table 3 of main text.}
    \label{tab:SMtable8}
    \begin{ruledtabular}
    \begin{tabular}{c c c c c}
    &\multicolumn{2}{c}{$J^{\mathrm{Co} \dotsb \mathrm{Co}}$} & \multicolumn{2}{c}{$J^{\mathrm {eff}}$}\\
          \cline{2-3}\cline{4-5}\\
         Structure &  S=3/2 &  S=5/2 &  S=3/2 &  S=5/2\\
          \hline
         $T_N^m$  (K)  & 49.60 & 41.66& 61.39 & 51.57      \\
         $|\Theta^m|$ (K)  & 95.77 & 80.45& 114.98 & 96.58    \\
    \end{tabular}
    \end{ruledtabular}
\end{table}

\bibliography{references}